\documentclass[11pt]{article}
\title{\bf New $q$-ary Quantum MDS Codes with Distances Bigger than $\frac{q}{2}$}
\author{Xianmang He, Liqing Xu and Hao Chen
  \thanks{X. He is with the School of Information Science and Technology, Ningbo University, Ningbo 315211, Zhejiang Province, China, hexianmang@nbu.edu.cn . L.Xu and H. Chen are with the Department of Mathematics, School of Sciences, Hangzhou Dianzi University, Hangzhou  310018, Zhejiang Province, China, lqxu@hdu.edu.cn, haochen@hdu.edu.cn. H.He was supported by NSFC Grant 31521101. L. Xu and H.Chen were supported by NSFC Grants 11371138 and 11531002.}}

\begin{document}

\maketitle
\begin{abstract}

Constructions of quantum MDS codes have been studied by many authors. We refer to the table in page 1482 of [3] for known constructions. However there have been constructed only a few  $q$-ary quantum MDS $[[n,n-2d+2,d]]_q$ codes with minimum distances $d>\frac{q}{2}$ for sparse lengths $n>q+1$. In the case $n=\frac{q^2-1}{m}$ where $m|q+1$ or $m|q-1$ there are complete results. In the case  $n=\frac{q^2-1}{m}$ while $m|q^2-1$ is not a factor of $q-1$ or $q+1$, there is no $q$-ary quantum MDS code with $d> \frac{q}{2}$ has been constructed. In this paper we propose a direct approach to construct Hermitian self-orthogonal codes over ${\bf F}_{q^2}$. Then we give some new $q$-ary quantum codes in this case. Moreover we present many new $q$-ary quantum MDS codes with lengths of the form $\frac{w(q^2-1)}{u}$ and minimum distances $d > \frac{q}{2}$. \\
\end{abstract}


\section{Introduction}

Quantum error-correcting codes are important for quantum information processing and quantum computation. The construction of quantum error-correcting codes has been an active field of quantum information theory since the publication of [19, 20, 15].  It is known for any pure quantum $[[n, k, d]]_q$ code the parameters satisfy the quantum Singleton bound $k \leq n-2d+2$. The $q$-ary quantum codes reaching this bound are called quantum MDS codes ([14, 15, 2]). Many constructions of $q$-ary quantum MDS codes have been proposed based on the Hermitian self-orthogonal codes over ${\bf F}_{q^2}$.\\

The Hermitian inner product over ${\bf F}_{q^2}^n$ is defined as follows. $<{\bf u},{\bf v}>_h=u_1v_1^q+\cdots+u_nv_n^q$, where ${\bf u}=(u_1,...,u_n)$ and ${\bf v}=(v_1,...,v_n)$ are vectors in ${\bf F}_{q^2}^n$. The following result gives a construction of $q$-ary quantum MDS  codes from Hermitian orthogonal codes over ${\bf F}_{q^2}$.\\

{\bf Theorem 1.1 ([2])} {\em If ${\bf C}$ is a $[n,k,n-k+1]_{q^2}$ MDS code over ${\bf F}_{q^2}$ which is orthogonal under the Hermitian inner product. Then we have a $q$-ary quantum MDS $[[n, n-2k, k+1]]_q$ code.}\\

There have been published many papers on the construction of quantum MDS  codes ([1, 2, 4, 5, 6, 7, 8, 9, 10, 11, 12, 13, 14, 15, 16, 17]). They were constructed  from generalized Reed-Solomon codes ([8, 9, 10]), cyclic or constacyclic codes ([7, 11, 12, 3]). However it seems that for many lengths $q+1<n< q^2-1$ if there is a $q$-ary quantum MDS  code with length $n$ and minimum distance $d>\frac{q}{2}$ is still an un-solved problem. For only very few sparse  lengths such $q$-ary quantum MDS codes with $d>\frac{q}{2}$ have been constructed ([7, 8, 9, 10, 11, 12, 3, 21, 22]). In the case length $n=\frac{q^2-1}{m}$ where $m$ is an integer satisfying $m|q+1$ or $m|q-1$  the following results have been proved ( [3, 13, 21], or see lines 13, 14 and 20 in the table of page 1482 of [3]). \\

1) For odd prime powers $q=2^es+1$ where $s$ is odd and odd number $\lambda |s$ and $f \leq e-1$, quantum MDS $[[2^f\lambda(q+1), 2^f\lambda(q+1)-2d+2, d]]_q$ codes with the minimum distance $d$ equal to any integer in the range $2 \leq \frac{q+1}{2}+2^f\lambda$ are constructed ([3] Theorem 4.11).\\

2) In the case $m|q+1$ and $m$ odd  there is a $q$-ary quantum MDS  code with length $\frac{q^2-1}{m}$ and any minimum distance $d$ in the range $2 \leq d \leq \frac{q+1}{2}+\frac{q+1}{2m}-1$. In the case  $m|q+1$  and $m$ even there is a $q$-ary quantum MDS  code with length $\frac{q^2-1}{m}$ and any minimum distance $d$ in the range $2 \leq d \leq \frac{q+1}{2}+\frac{q+1}{m}-1$ ( see [3, 21]).\\

However in the case  $n=\frac{q^2-1}{m}$ where $m|q^2-1$ is not a factor of $q-1$ or $q+1$ no $q$-ary quantum MDS code with length $\frac{q^2-1}{m}$ and minimum distance $d > \frac{q}{2}$ has been constructed. Though in this case each cyclotomic set has only one element the technique in [8, 12, 13, 3] is not sufficient to get the desirable $q$-ary quantum codes. In this paper some new $q$-ary quantum MDS codes in this case with minimum distance $d>\frac{q}{2}$ are constructed. We use a direct approach of constructing generator matrices of Hermitian orthogonal MDS codes over ${\bf F}_{q^2}$.  Many new $q$-ary quantum MDS  codes for the length $n=\frac{w(q^2-1)}{u}$ and $d>\frac{q}{2}$ for some integers $w$ and $u$ are  also presented.\\

We need the following Lemmas in this paper.\\

{\bf Lemma 1.1.} {\em If $\theta$ is a primitive element of the multiplicative group ${\bf F}_{q^2}^{*}$. Suppose $m$ is a factor of $q^2-1$. Then $\Sigma_{j=1}^{\frac{q^2-1}{m}} \theta^{jtm}=0$ except the case that $t$ can be divisible by $\frac{q^2-1}{m}$.}\\

{\bf Proof.} For any $1 \leq t \leq \frac{q^2-1}{m}-1$, $\theta^{mt}$ generates a subgroup $G$ of ${\bf Z}/\frac{q^2-1}{m}{\bf Z}$ which is generated by $\theta^{m}$.  The order of the group $G$ is $\frac{\frac{q^2-1}{m}}{gcd(t, \frac{q^2-1}{m})} >1$. Since $G \neq \{1\}$, for any non-unit element $\theta^{mt}$, $\theta^{mt} G=G$. Thus $\theta^{mt} \Sigma_{j=1}^{\frac{q^2-1}{m}} \theta^{mtj} = \Sigma_{j=1}^{\frac{q^2-1}{m}} \theta^{mtj}$. It is clear $\theta^{mt} \neq 1$, the conclusion follows directly.\\

{\bf Lemma 1.2.} {\em Suppose $v_0, ...,v_n$ are $n$ non-zero elements in the multiplicative group ${\bf F}_q^*$. If ${\bf g}_l=(g_{1l},...,g_{nl})$ where $l=1,...,k$, are $k$ linear independent rows  in ${\bf F}_{q^2}^n$ satisfying that $\Sigma_{j=1}^n v_j g_{jl_1} g_{jl_2}^q=0$ for any two indices $l_1$ and $l_2$ in the set $\{1,...,k\}$ (here $l_1=l_2$ is possible). Then we have a Hermitian orthogonal $[n, k]_{q^2}$ code generated by these $k$ rows.}\\

{\bf Proof.} We can set $v_j=(v_j')^{q+1}$ for $j=1,...,n$. Thus the equivalent code $(v_1',...,v_n') {\bf C}$ is a Hermitian orthogonal code, where ${\bf C}$ is a $q^2$-ary code generated by these $k$ rows ${\bf g}_1,...,{\bf g}_k$.\\

\section{New Quantum MDS Codes I}

{\bf Construction 1.} Let $m$ be a factor of $q^2-1$. For any fixed positive integer $w$ we define a length $\frac{q^2-1}{m}$ linear error codes over ${\bf F}_{q^2}$ as follows. \\

$$
\begin{array}{ccccc}
{\bf C}_w=\{(\theta^{m} f(\theta^{m}), \theta^{2m}f(\theta^{2m}),...,\theta^{jm}f(\theta^{jm}),...,\\ \theta^{(\frac{q^2}{m}-1)m} f(\theta^{(\frac{q^2}{m}-1)m}), f(1)): f \in {\bf F}_{q^2}[x],deg(f) \leq w-1\}\\
\end{array}
$$

 It is clear that ${\bf C}_w$ is a MDS $[\frac{q^2-1}{m}, w, \frac{q^2-1}{m}-w+1]$ code over ${\bf F}_{q^2}$. The Hermitian inner product of  any two codewords (corresponding to two polynomials $f$ and $g$) is $\Sigma_{j=1}^{\frac{q^2-1}{m}} \theta^{jm+jqm} fg^q(\theta^{jm})$. Thus we only need to check $\Sigma_{j=1}^{\frac{q^2-1}{m}} \theta^{(q+1)jm} \theta^{t_1jm} \theta^{t_2jqm}=\Sigma_{j=1}^{\frac{q^2-1}{m}} \theta^{(q+1)mj} \theta^{jm(t_1+t_2q)}=\Sigma_{j=1}^{\frac{q^2-1}{m}} \theta^{jm(q+1+t_1+t_2q)}$, where $0 \leq t_1, t_2 \leq w-1$.\\

{\bf Theorem 2.1.} {\em Let $m$ be a factor of $q+1$. If for all non-negative integers $t_1$ and $t_2$ satisfying $0 \leq t_1, t_2 \leq w-1$, $q+1+t_1+t_2q$ cannot be divisible by $\frac{q^2-1}{m}$. The code ${\bf C}_w$ is Hermitian self-orthogonal. If $m=2k+1$ is an odd positive integer and $w < \frac{k+1}{2k+1}(q-1)$. The above condition is satisfied.}\\

{\bf Proof.} The first conclusion is obvious. It is sufficient to prove that  if $w <\frac{k+1}{2k+1}(q-1)$, $q+1+t_1+t_2q$, where  $t_1<w,t_2<w$, cannot be divisible $\frac{q^2-1}{m}$. Since $q+1+t_1+t_2q \leq (q+1)(1+w-1) <(k+1)\frac{q^2-1}{m}$,  we have  if $q+1+t_1+t_2q$ can be divisible by $\frac{q^2-1}{m}$, the quotient $\frac{q+1+t_1+t_2q}{\frac{q^2-1}{m}}  \leq k$. On the other hand $\frac{q^2-1}{m}=\frac{q+1}{m}q-\frac{q+1}{m}$. That is $\frac{q^2-1}{m} \equiv q-\frac{q+1}{m}$ $mod$ $q$ because $\frac{q+1}{m}$ is an integer. Therefore if $q+1+t_1+t_2q$ can be divisible by $\frac{q^2-1}{m}$,  then residue of $q+1+t_1+t_2q$ module $q$ is in the range $[\frac{k+1}{m}(q+1)-1, q-1]$. It is obvious that the residue of $q+t_2q+1+t_1$ module $q$ is $1+t_1 \leq w <\frac{k+1}{2k+1}(q-1)$. Since $\frac{k+1}{m}(q+1)$ is a positive integer and $\frac{k+1}{m}(q-1)=\frac{k+1}{m}(q+1)-1-\frac{1}{m}<\frac{k+1}{m}(q+1)$, the conclusion follows directly.\\

{\bf Corollary 2.1.} {\em If $m=2k+1$ is an odd factor of $q+1$, for each positive integer $d$ in the range $2\leq d \leq [\frac{k+1}{2k+1}(q-1)+1]$, there exists a $q$-ary quantum MDS code with length $\frac{q^2-1}{m}$ and minimum distance $d$.}\\

Suppose $q$ is a prime power and $q+1=\lambda r$ where $r$ is odd, then for each $d$ in the range $2 \leq d \leq \frac{q-1}{2}+\frac{\lambda}{2}$, length $\lambda(q-1)$ $q$-ary quantum MDS code with the minimum distance $d$ were constructed in [3, 12, 13, 21]. Their construction was based on constacyclic codes over ${\bf F}_{q^2}$. However this kind of quantum $q$-ary MDS codes can be constructed from Corollary 2.1 directly. From the construction 1, we can give the generator matrix of the corresponding  MDS Hermitian self-orthogonal codes over ${\bf F}_{q^2}$ immediately.\\

The construction 1 can be extended to $[\frac{q^2-1}{m}+1, w+1, \frac{q^2-1}{m}-w+1]$ Hermitian self-orthogonal code over ${\bf F}_{q^2}$ with the following generator matrix.\\

$$
\left(
\begin{array}{ccccccccccccccc}
\frac{q+1}{m}&1&\cdots&1&1\\
0&\theta^{m}&\cdots&\theta^{(\frac{q^2-1}{m}-2)m}&\theta^{\frac{q^2-1}{m}m}=1\\
\cdots&\cdots&\cdots&\cdots&\cdots\\
0&\theta^{im}&\cdots&\theta^{(\frac{q^2-1}{m}-2)im}&1\\
\cdots&\cdots&\cdots&\cdots&\cdots\\
0&\theta^{wm}&\cdots&\theta^{(\frac{q^2-1}{m}-2)wm}&1\\
\end{array}
\right)
$$

Therefore we have the following result which can be thought as a generalization of Theorem 4.4 of [11].\\

{\bf Theorem 2.2.} {\em For each odd number $m=2k+1$ satisfying $m|q+1$, we have $[[\frac{q^2+m-1}{m}, \frac{q^2+m-1}{m}-2d,d+1]]_q$ quantum MDS code for each $d$ in the range $2 \leq d \leq [\frac{k+1}{2k+1}(q-1)+1]$.}\\

In the case $q+1$ can be divisible by $3$, we have length $\frac{q^2-1}{3}+1=\frac{q^2+2}{3}$ $q$-ary quantum MDS code with minimum distance $d$ for each $d$ in the range $2 \leq d \leq \frac{2(q+1)}{3}$. This recovers the 2nd conclusion of Theorem 4.4 of [10]. Moreover if $5|q+1$, then we have length $\frac{q^2-1}{5}+1=\frac{q^2+4}{5}$ $q$-ary quantum MDS code with the minimum distance $d$ for each $d$ in the range $2 \leq d \leq \frac{3(q+1)}{5}$ and many other new quantum MDS codes.\\

\begin{center}
{\bf Table 1} $[[\frac{q^2+m-1}{m}, \frac{q^2+m-1}{m}-2d, d+1]]_q$ quantum MDS codes\\
\bigskip
\begin{tabular}{||c|c||}\hline
Quantum MDS code& $q, m,d $ \\ \hline
$[[33,15,10]]_{17}$&17, 9, 9\\ \hline
$[[73,51,12]]_{19}$&19, 5, 11\\ \hline
$[[57, 27, 16]]_{29}$&29, 15, 15\\ \hline
$[[73, 35, 20]]_{37}$&37, 19, 19\\ \hline
$[[81, 41, 21]]_{41}$&41, 21, 20\\ \hline
$[[169, 125, 23]]_{43}$&43,11, 22\\ \hline
$[[105, 53, 27]]_{53}$&53, 27, 26\\ \hline
\end{tabular}
\end{center}

We need the following two Lemmas in construction 2.\\

{\bf Lemma 2.1.} {\em Suppose $q$ is an even prime power $2^h$. Let $\theta \in {\bf F}_{q^2}$ be a primitive element which generate the multiplicative group ${\bf F}_{q^2}^{*}$. If $m_1$  and $m_2$ are factors of $q^2-1$ satisfying $gcd(m_1,m_2)=1$. We set $m_3=\frac{q^2-1}{m_1}$ and $m_4=\frac{q^2-1}{m_2}$. Let ${\bf M}_1$  be the set of all indices $j$ satisfying $1 \leq j \leq m_3-1$ and $j$ cannot be divisible by $m_2$, and ${\bf M}_2$ be the set of all indices $j$ satisfying $1 \leq j \leq m_4-1$ and $j$ cannot be divisible by $m_1$. Then $\Sigma_{j \in {\bf M}_1} \theta^{m_1tj}+\Sigma_{j \in {\bf M}_2} \theta^{m_2tj}=0$ for $t=1,...,min\{m_3,m_4\}-1$.}\\

{\bf Proof.} From Lemma 1.1 and the fact $-1=1$ in the finite field ${\bf F}_{2^{2h}}$ we get the conclusion. Here we should note that in the two equalities $\Sigma_{j=1}^{m_3} \theta^{m_1tj}=0$ and $\Sigma_{j=1}^{m_4} \theta^{m_2tj}=0$. The common part is $\Sigma_{j=1}^{m_5} \theta^{m_1m_2tj}$, where $m_5=\frac{q^2-1}{m_1m_2}$.\\

Here $|{\bf M}_1|=m_3-\frac{q^2-1}{m_1m_2}$ and $|{\bf M}_2|=m_4-\frac{q^2-1}{m_1m_2}$.\\

Similarly we have the following result.\\

{\bf Lemma 2.2.} {\em Suppose $q$ is an even prime power $2^h$. Let $\theta \in {\bf F}_{q^2}$ be a primitive element which generate the multiplicative group ${\bf F}_{q^2}^{*}$. If $m_1,...,m_s$ are factors of $q^2-1$ satisfying $gcd(m_{s_1},m_{s_2})=1$ for any $s_1 \neq s_2$. We set $m_1'=\frac{q^2-1}{m_1},..., m_s'=\frac{q^2-1}{m_s}$. Set ${\bf M}_u$ the subgroup of the multiplicative group ${\bf F}_{q^2}^*$ generated by $\theta^{m_u}$. Let ${\bf M}_{s_1,...s_l}$  be the intersection of ${\bf M}_{s_1},...,{\bf M}_{s_l}$ for distinct indices $s_1, ..., s_l$ in the set $\{1,...,s\}$. The set ${\bf M}$ is the set ${\bf M}_1 \cup \cdots \cup {\bf M}_s$ deleting these elements in ${\bf M}_{s_1,...,s_l}$  where $l$ is even and remaining these elements in ${\bf M}_{s_1,...,s_{l'}}$ where $l'$ is odd. Then $\Sigma_{j \in {\bf M} \cap {\bf M}_1} \theta^{m_1tj}+\Sigma_{j \in {\bf M}_2 \cap {\bf M}} \theta^{jm_2t}+\cdots+\Sigma_{j \in {\bf M} \cap {\bf M}_s} \theta^{jm_2t}=0$ for $t=1,...,min\{m_1',...,m_s' \}-1$.}\\

Since the characteristic of the field ${\bf F}^{2^{2h}}$ is $2$, the sum of even number of a same element in this field is always zero, and  the sum of odd number of a same element in this field is this element. Thus in the summation $\Sigma_{j \in {\bf M} \cap {\bf M}_1} \theta^{m_1tj}+\Sigma_{j \in {\bf M}_2 \cap {\bf M}} \theta^{jm_2t}+\cdots+\Sigma_{j \in {\bf M} \cap {\bf M}_s} \theta^{jm_2t}$, for these elements $\theta^{m_{s_1}\cdots m_{s_{l'}}t'}$  in ${\bf M}$ which are in ${\bf M}_{s_1,...,s_{l'}}$, $l'$ is an odd number, there is only one $\theta^{m_{s_1}\cdots m_{s_{l'}}t't}$ in the above summation.\\

{\bf Construction 2.} If $q$ be an even prime power $2^h$, $m_1$ and $m_2$ are odd positive integers satisfying $gcd(m_1,m_2)=1, m_1|q+1, m_2|q+1$.  We set $m_3=\frac{q^2-1}{m_1}$, $m_4=\frac{q^2-1}{m_2}$, $M=m_3+m_4-\frac{2(q^2-1)}{m_1m_2}$. We construct a length $M$ linear code ${\bf C_M}$ over ${\bf F}_{q^2}$ as follows. For each polynomial $f \in {\bf F}_{q^2}[x]$ with degree less than or equal to $w-1$, the codeword is a length $M$ vector with the coordinate at the element $u \in {\bf M}_1 \cup {\bf M}_2-{\bf M}_1 \cap {\bf M}_2$,  $\theta^{m_1j}f(\theta^{m_1j})$ (if $u=\theta^{m_1j} \in {\bf M}_1$), or $\theta^{m_2j} f(\theta^{m_2j})$ (if $u=\theta^{m_2j} \in {\bf M}_2$). It is clear ${\bf C_M}$ is a $[M, w, M-w+1]$ MDS code over ${\bf F}_{q^2}$. We need to check the exponential sum $\Sigma_{j \in {\bf M}_1} \theta^{jm_1(q+1+t_1+t_2q)}+\Sigma_{j \in {\bf M}_2} \theta^{jm_2(q+1+t_1+t_2q)}$.\\

{\bf Theorem 2.3.} {\em Let $m_1, m_2, m_3, m_4, M$ and $w$ be positive integers as above. If for all non-negative integers $t_1$ and $t_2$ satisfying $0 \leq t_1, t_2 \leq w-1$, $q+1+t_1+t_2q$ cannot be divisible by $m_3$ and $m_4$. The code ${\bf C_M}$ is Hermitian self-orthogonal. If $m_1=2k_1+1 < m_2=2k_2+1$ are odd positive integers and $w < \frac{k_2+1}{2k_2+1}(q-1)$. The above condition is satisfied.}\\

{\bf Proof.} The conclusion follows from the proof of Theorem 2.1 and the fact $w < min\{\frac{k_1+1}{2k_1+1}(q-1), \frac{k_2+1}{2k_2+1}(q-1)\}$.\\

{\bf Corollary 2.2.} {\em Suppose that $q$ is an even prime power $2^h$, $m_1=2k_1+1$ and $m_2=2k_2+1$ are odd positive integers satisfying $gcd(m_1,m_2)=1, m_1<m_2$ and $m_1|q+1, m_2|q+1$. We set $m_3=\frac{q^2-1}{m_1}$, $m_4=\frac{q^2-1}{m_2}$, $M=m_3+m_4-\frac{2(q^2-1)}{m_1m_2}$. For each positive integer $d$ in the range $2 \leq d \leq [\frac{k_2+1}{2k_2+1}(q-1)+1]$, there is a length $M$ $q$-ary quantum MDS code with the minimum distance $d$.}\\

From Lemma 2.2 we can generalize to the case that $q+1$ has several factors $m_1,...m_s$, where $gcd(m_{s_1}, m_{s_2})=1$ for $s_1 \neq s_2$.\\

Let $q$ be an even prime power $2^h$, suppose $q+1=m_1m_2$ where $m_1$ and $m_2$ are two coprime factors of $q+1$ satisfying $2k_1+1=m_1 <m_2=2k_2+1$. From Corollary 2.2 we have a length $m_2+m_1-2)(q-1)$ $q$-ary quantum MDS code with the minimum distance $d$ for each $d$ in the range $ 2\leq d \leq [\frac{k_2+1}{2k_2+1}(q-1)+1]$. For example, for $h=5$, we have $32$-ary  quantum MDS code. Since $372=12 \cdot (2^5-1)$ where $12$ is not a factor of $2^5+1$, this quantum MDS code is not covered in the previous constructions. Similarly we have $[[1008, 942, 33]]_{64}$ quantum MDS code and more such codes for $h=5,6,7,9....$ as in the following table.\\

\begin{center}
{\bf Table 2} $[[(m_1+m_2-2)(2^h-1), (m_1+m_2-2)(2^h-1)-2k, k+1]]_{2^h}$ quantum MDS codes\\
\bigskip
\begin{tabular}{||c|c||}\hline
Quantum MDS code& $h, m_1,m_2, k $ \\ \hline
$[[372,340,17]]_{32}$&5, 3,11, 16\\ \hline
$[[1008,942,34]]_{64}$&6, 5, 13, 32\\ \hline
$[[5588, 5460, 65]]_{128}$&7, 3, 43, 64\\ \hline
$[[22484, 21956, 265]]_{512}$&9, 19, 27, 264\\ \hline
\end{tabular}
\end{center}

Let $q$ be an even prime power $2^h$, suppose $q+1=m_1m_2m_3$ where $m_1, m_2$ and $m_3$ are three factors of $q+1$ satisfying $m_1 <m_2<m_3=2k_3+1$. Suppose that any two of $m_1,m_2,m_3$ are coprime. Then the size of the set ${\bf M}$ in Lemma 2.3 is $|{\bf M}|=(q^2-1)(\frac{1}{m_1}+\frac{1}{m_2}+\frac{1}{m_3}-\frac{2}{m_1m_3}-\frac{2}{m_1m_3}-\frac{2}{m_2m_3}+\frac{4}{m_1m_2m_3})$. For each $d$ in the range $2 \leq d \leq [\frac{k_3+1}{2k_3+1}(q-1)+1]$ we have a length $|{\bf M}|$ $q$-ary quantum MDS code with the minimum distance $d$.\\

Actually in the case $q$ is an odd prime power, we can use the equivalent codes to get new quantum MDS codes as follows. If $q$ is an odd prime power, then $2$ is a non-zero element in ${\bf F}_q \subset {\bf F}_{q^2}$. If $m_1=2k_1+1<m_2=2k_2+1$ are two odd factors of $q+1$. Then we have the following identity. If $t$ cannot be divisible by $\frac{q^2-1}{m_1}$ or $\frac{q^2-1}{m_2}$, then

$$
\begin{array}{ccccccc}
\Sigma_{j=1}^{\frac{q^2-1}{m_1}} \theta^{m_1tj}+\Sigma_{j=1}^{\frac{q^2-1}{m_2}} \theta^{m_2jt}=0
\end{array}
$$

For these indices $j$'s which are in both parts, that is, $j=m_1m_2j'$, we have $2 \theta^{m_1m_2 tj'}$ in the above identity. Since $2=u^{q+1}$ for some $u \in {\bf F}_{q^2}$. The equivalent codes can be used to get Hermitian orthogonal codes from Lemma 1.2.  Thus we have the following result.\\

{\bf Theorem 2.3.} {\em Suppose that $q$ is an odd prime power, $m_1=2k_1+1$ and $m_2=2k_2+1$ are odd positive integers satisfying $gcd(m_1,m_2)=1, m_1<m_2$ and $m_1|q+1, m_2|q+1$. We set $m_3=\frac{q^2-1}{m_1}$, $m_4=\frac{q^2-1}{m_2}$, $M=m_3+m_4-\frac{q^2-1}{m_1m_2}$. For each positive integer $d$ in the range $2 \leq d \leq [\frac{k_2+1}{2k_2+1}(q-1)+1]$, there is a length $M$ $q$-ary quantum MDS code with the minimum distance $d$.}\\

In the following table we give some new quantum MDS $q$-ary codes from Theorem 2.3. \\

\begin{center}
{\bf Table 3} $[[\frac{q^2-1}{m_1}+\frac{q^2-1}{m_2}-\frac{q^2-1}{m_1m_2}, \frac{q^2-1}{m_1}+\frac{q^2-1}{m_2}-\frac{q^2-1}{m_1m_2}-2k, k+1]]_{q}$ quantum MDS codes\\
\bigskip
\begin{tabular}{||c|c||}\hline
Quantum MDS code& $q, m_1,m_2, k $ \\ \hline
$[[412,412-2k,k+1]]_{29}$&29, 3,5, $1 \leq k \leq 16$\\ \hline
$[[720,720-2k,k+1]]_{41}$&41, 3, 7, $1\leq k \leq 22$\\ \hline
$[[1624, 1624-2k, k+1]]_{59}$&59, 3, 5, $1 \leq k \leq 34$\\ \hline
$[[2952, 2952-2k, k+1]]_{83}$&83, 3,7, $1\leq k \leq 46$\\ \hline
\end{tabular}
\end{center}

\section{New Quantum Codes  II}

\subsection{ Odd $q$ and even $m|q-1$ (Recovery of Theorem 4.11 in [3])}

Suppose $q$ is an odd prime power and $q-1=2^ha$ where $a$ is an odd number. We assume $m=2^{h_1}a_1 \geq 6$ is an even factor of $q-1$ where $a_1$ is an odd number. Then $h_1 \leq h$ and $a_1$ is a factor of $a$ and  $a=a_1a_2$ for a positive integer $a_2$.  We first prove the following identity holds when $0 \leq t_1, t_2 \leq \frac{q+1}{2}+2^{h-h_1}a_2-2$.\\

$$
\begin{array}{cccccc}
\Sigma_{j=1}^{\frac{q^2-1}{m}} \theta^{jm(t_1+t_2q+\frac{q+1}{2})}=0
\end{array}
$$

From the condition $m \geq 6$, $t_1+t_2q+\frac{q+1}{2}<q^2-1$. Thus if $(t_1+\frac{q+1}{2})+t_2q$ can be divisible by $\frac{q^2-1}{m}$. The quotient $u < m$. In the case $t_1+\frac{q+1}{2} \leq q-1$, we  have $u\frac{q^2-1}{m}=t_2q+t_1+\frac{q+1}{2}$. The quotient is $t_2$ and the remainder is $t_1+\frac{q+1}{2}$. The quotient and the remainder have to be the same since $u(\frac{q-1}{m})$ is an integer.\\

Since $t_1+\frac{q+1}{2}=t_2$ can be divisible by $\frac{q-1}{m}$,  $t_1+1+\frac{q-1}{2}$  can be  divisible by $\frac{q-1}{m}=2^{h-h_1}a_2$. From $t_1 \geq 0$ we have $t_1+1 \geq 1$, and $t_1 \geq 2^{h-h_1}a_2-1$. On the other hand $t_2=t_1+\frac{q+1}{2}$, $t_2 \geq \frac{q+1}{2}+2^{h-h_1}a_2-1$. This is a contradiction. Thus $t_1+t_2q+\frac{q^2-1}{2^{h-h_1+1}m}$ can not be divisible by $\frac{q^2-1}{m}$. \\

In the case $t_1+\frac{q+1}{2} \geq q$. The $u\frac{q^2-1}{m}=(t_2+1)q+(t_1-\frac{q-1}{2})$. The quotient is $t_2+1$ and the remainder is $t_1-\frac{q-1}{2}$. These two numbers have to be the same since $u< m$. Thus $t_2+1=t_1-\frac{q-1}{2}$ can be divisible by $\frac{q-1}{m}=2^{h-h_1}a_2$. From $t_2+1 \geq 1$, we have $t_2 \geq 2^{h-h_1}a_2-1$. Thus $t_1 \geq t_2+1+\frac{q-1}{2} \geq \frac{q+1}{2}+2^{h-h_1}a_2-1$. This is a contradiction.\\

We can set $v_j'=\theta^{j\frac{m(q+1)}{2}} \in {\bf F}_q^{*}$. ${\bf g}_l=(\theta^{ml}, \theta^{2ml},...,\theta^{jml},...,\theta^{\frac{q^2-1}{m}ml})$, where $0 \leq l \leq \frac{q+1}{2}+(2^{h-h_1+1}-1)a_3+2^{h-h_1}-2$. Thus  a $[\frac{q^2-1}{m}, k]_{q^2}$ Hermitian orthogonal code can be constructed from Lemma 1.1. This is actually equivalent to the evaluation code of all polynomials of the form $xf(x)$'s, where $f(x)$ is a polynomial with degree $deg(f) \leq \frac{q+1}{2}+(2^{h-h_1+1}-1)a_3+2^{h-h_1}-3$. It is a MDS code. From Theorem 1.1 we have length $\frac{q^2-1}{m}$ quantum MDS $q$-ary code with the minimum distance $d=k+1$ in the range $2 \leq d \leq \frac{q+1}{2}+2^{h-h_1}a_2$.\\

{\bf Theorem 3.1.} {\em If $q=2^ha+1$ is an odd prime power where $a$ is an odd number and $m=2^{h_1}a_1 \geq 6$ is an even factor of $q-1$ where $a_1|a$ is an odd factor of $a$. Then we have $q$-ary quantum MDS codes with  length $\frac{q^2-1}{m}$ and any minimum distance $d$ in the range $2 \leq d \leq \frac{q+1}{2}+2^{h-h_1}a_2$.}\\

Actually Theorem 3.1 recovers Theorem 4.11 in [3].\\

\subsection{Length $\frac{w(q^2-1)}{u}$ quantum $q$-ary MDS codes}

Suppose $m_1=2^{h_1}a_1 \geq 6$ and $m_2=2^{h_2}b_1 \geq 6$ are two even factors of $q-1=2^ha_1a_2=2^hb_1b_2$ where $a_1,a_2,b_1,b_2$ are odd numbers.  Then we have two identities of the form in the previous subsection. The addition of these two identities gives another identity. For these indices $j$ which can be divisible by $m_1$ and $m_2$, we have to use the element $\theta^{j \frac{m_1(q+1)}{2}}+\theta^{j\frac{m_2(q+1)}{2}} \in {\bf F}_q$. It is obvious this is a non-zero element in ${\bf F}_q^{*}$ when $lcm(m_1,m_2)=q-1$ (here $lcm$ is the least common multiple). Set ${\bf M}_1$ the set of indices $m_1 \cdot \{1,...,\frac{q^2-1}{m_1}\}$ and ${\bf M}_2=m_2 \cdot \{1,...,\frac{q^2-1}{m_2}\}$, ${\bf M}={\bf M}_1 \cup {\bf M}_2$. Here  $|{\bf M}|=|{\bf M}_1|+|{\bf M}_2|-\frac{q-1}{lcm(m_1,m_2)}(q+1)=\frac{q^2-1}{m_1}+\frac{q^2-1}{m_2}-(q+1)$ when $lcm(m_1, m_2)=q-1$.\\

{\bf Theorem 3.2.} {\em If $q=2^{h_1}a_1a_2+1=2^{h_2}b_1b_2+1$ is an odd prime power as above and $a_1, a_2, b_1, b_2$ are odd numbers. $m_1=2^{h_1}a_1$ and $m_2=2^{h_2}b_1$ are two even factors of $q-1$ satisfying $lcm(m_1,m_2)=q-1$ as above. Then we have $q$-ary quantum MDS codes with  length $|{\bf M}|=\frac{q^2-1}{m_1}+\frac{q^2-1}{m_2}-(q+1)$ and any minimum distance $d$  in the range $2 \leq d \leq \frac{q+1}{2}+min\{2^{h-h_1}a_2, 2^{h-h_2}b_2\}$.}\\

{\bf Corollary 3.1.} {\em If $2m_1m_2+1$ is a prime power where $m_1<m_2$ are two co-prime odd numbers. Then we have a length $\frac{(m_1+m_2-1)(q^2-1)}{2m_1m_2}=(m_1+m_2-1)(2m_1m_2+2)$ $q$-ary quantum MDS $[[(m_1+m_2-1)(2m_1m_2+2), (m_1+m_2-1)(2m_1m_2+2)-2d+2, d]]_q$ code with any $d$ in the range $2 \leq d \leq m_1m_2+m_1+1$.}\\

From Corollary 3.1 we list some new quantum MDS codes in the following table.\\

\begin{center}
{\bf Table 4} $[[(m_1+m_2-1)(2m_1m_2+2), (m_1+m_2-1)(2m_1m_2+2)-2d+2, d]]_{q}$ quantum MDS codes\\
\bigskip
\begin{tabular}{||c|c||}\hline
Quantum MDS code& $q, m_1,m_2, d $ \\ \hline
$[[224,224-2d+2,d]]_{29}$&31, 3, 5, $2 \leq d \leq 17$\\ \hline
$[[396,396-2d+2,d]]_{41}$&43, 3, 7, $2\leq d \leq 25$\\ \hline
$[[884, 884-2d+2, d]]_{67}$&67, 3, 11, $2 \leq d \leq 37$\\ \hline
$[[792, 792-2d+2, d]]_{71}$&71, 5, 7, $2 \leq d \leq 41$\\ \hline
$[[1196, 1196-2d+2,d]]_{91}$&91, 5, 9, $2 \leq d \leq 51$\\ \hline
\end{tabular}
\end{center}

The method  can be extended to the more general case. Let $q=2^ha_1^1a_2^1+1=2^ha_1^2a_2^2+1=\cdots=2a_1^sa_2^s+1$ be an odd prime power and $a_i^j$'s are odd numbers. If $m_i=2^{h_i}a_1^i \geq 6$ are even factors of $q-1$ where $h_i \leq h$ and $i=1,...,s$. We set ${\bf A}=min\{2^{h-h_1}a_2^1,...,2^{h-h_j}a_2^j,...,2^{h-h_s}a_2^s \}$, ${\bf M}_j=m_j \cdot \{1,..,\frac{q^2-1}{m_j}\}$ for $j=1,...,s$ and ${\bf M}={\bf M}_1 \cup \cdots \cup {\bf M}_s$.\\

{\bf Theorem 3.3.} {\em Suppose $\theta^{j\frac{m_{i_1}(q+1)}{2}}+\cdots+\theta^{j\frac{m_{i_t}(q+1)}{2}} \neq 0$ for these indices $1 \leq j \leq max\{\frac{q^2-1}{2m_{i_1}},..,\frac{q^2-1}{2m_{i_t}}\}$ and $j$ can be divisible by $m_{i_1},...,m_{i_t}$. There is a $q$-ary quantum MDS code with the length $|{\bf M}|$ and any minimum distance $d$  in the range $2 \leq d \leq \frac{q+1}{2}+A$.} \\

{\bf Corollary 3.2.} {\em If $q=2m_1\cdots m_s+1$ is an odd prime power and $m_1<m_2 \cdots <m_s$ are odd numbers and $\theta^{j\frac{m_{i_1}(q+1)}{2}}+\cdots+\theta^{j\frac{m_{i_t}(q+1)}{2}} \neq 0$ for  $1 \leq j \leq max\{\frac{q^2-1}{2m_{i_1}},..,\frac{q^2-1}{2m_{i_t}}\}$ satisfying that $j$ can be divisible by $m_{i_1},...,m_{i_t}$. Suppose $gcd(m_{i_1},m_{i_2})=1$ for two distinct indices $i_1$ and $i_2$. There is a  $q$-ary quantum MDS code with the length $|{\bf M}|$ and any minimum distance $d$ in the range $2 \leq d \leq \frac{q+1}{2}+m_1 \cdots m_{s-1}$.}\\

The cardinality of the set $|{\bf M}|$ can be computed easily from the algebra of sets. In the following table  we give some new quantum codes from Corollary 3.2 in the case $s=3$. In this case $|{\bf M}|=(\frac{q-1}{2m_1}+\frac{q-1}{2m_2}+\frac{q-1}{2m_3}-\frac{q-1}{2m_1m_2}-\frac{q-1}{2m_1m_3}-\frac{q-1}{2m_2m_3}+1)(q+1)=\frac{m_1m_2+m_2m_3+m_1m_3-m_1-m_2-m_3+1}{2m_1m_2m_3} \cdot (q^2-1)$. The condition in Theorem 3.3 when $s=3$ is satisfied automatically.\\

\begin{center}
{\bf Table 5} Quantum MDS codes from Corollary 2.2 when $s=3$\\
\bigskip
\begin{tabular}{||c|c||}\hline
Quantum MDS code& $q, m_1,m_2, m_3, d $ \\ \hline
$[[12084,12084-2d+2,d]]_{211}$&211, 3, 5, 7, $2 \leq d \leq 121$\\ \hline
$[[28552, 28552-2d+2, d]]_{331}$&331, 3, 5, 11, $2 \leq d \leq 181$\\ \hline
$[[77736, 77736-2d+2, d]]_{631}$&631, 5, 7, 9, $2 \leq d \leq 351$\\ \hline
$[[80652, 80652-2d+2,d]]_{571}$&571, 3, 5, 19, $2 \leq d \leq 301$\\ \hline
\end{tabular}
\end{center}

\section{New Quantum Codes III}

Let $q$ be an odd prime power and $m_1=2k_1+1$ is an odd factor of $q+1$. From Theorem 2.1 we have the following identity holds when $0 \leq t_1,t_2 \leq \frac{q-1}{2}+\frac{q+1}{2m_1}-2$.\\

$$
\begin{array}{ccccccc}
\Sigma_{j=1}^{\frac{q^2-1}{m_1}} \theta^{jm_1(t_1+t_2q)} \cdot \theta^{jm_1(q+1)}=0
\end{array}
$$

From section 3 if $m_2|q-1$ is an even factor of $q-1$ we have the following identity when $0 \leq t_1, t_2 \leq \frac{q-1}{2}+\frac{q-1}{m_2}-1$.\\

$$
\begin{array}{cccccc}
\Sigma_{j=1}^{\frac{q^2-1}{m_2}} \theta^{jm_2(t_1+t_2q)} \cdot \theta^{j\frac{m_2(q+1)}{2}}=0
\end{array}
$$

We can get the following identity

$$
\begin{array}{ccccccc}
\Sigma_{j=1}^{\frac{q^2-1}{m_1}} \theta^{jm_1(t_1+t_2q)} \cdot \theta^{jm_1(q+1)}+ H(\Sigma_{j=1}^{\frac{q^2-1}{m_2}} \theta^{jm_2(t_1+t_2q)} \cdot \theta^{j\frac{m_2(q+1)}{2}})=0
\end{array}
$$
Here $H$ can be  any non-zero $H \in {\bf F}_q^{*}$ and  the common $t_1$ and $t_2$ are in the range $0 \leq t_1, t_2 \leq \frac{q-1}{2}+ min \{\frac{q+1}{2m_1}-2, \frac{q-1}{m_2}-1\}$. At the position $\theta^{m_1m_2t}$, it is clear that $\theta^{m_1^2m_2t(q+1)}+H\theta^{\frac{m_1m_2^2t(q+1)}{2}}$ is an element in ${\bf F}_q$. Since $\theta^{(m_1-\frac{m_2}{2})m_1m_2t(q+1)}$ can only be the $\frac{q-1}{m_2}$ non-zero elements in the sub-group of ${\bf F}_q^{*}$ generated by $\theta^{m_2(q+1)}$, there exists a $H \in {\bf F}_q^{*}$ such that $\theta^{m_1^2m_2t(q+1)}+H\theta^{\frac{m_1m_2^2t(q+1)}{2}}$ is a non-zero element in ${\bf F}_q^{*}$ for any possible $t$. Thus we have the following result.\\

{\bf Theorem 4.1.} {\em If $q$ is an odd prime power and $m_1$ is an odd factor of $q+1$ and $m_2$ an even factor of $q-1$. Then we have a $q$-ary quantum MDS code with length $\frac{q^2-1}{m_1}+\frac{q^2-1}{m_2}-\frac{q^2-1}{m_1m_2}$ and any minimum distance $d$  in the range $2 \leq d \leq \frac{q-1}{2}+min \{\frac{q+1}{2m_1}, \frac{q-1}{m_2}+1\}$.}\\

Actually Theorem 4.1 is quite general as illustrated in the following results.\\

{\bf Corollary 4.1.} {\em Let $q$ be an odd prime power. If there exists an odd $m | q+1$ such that $m-1$ is an even factor of $q-1$. Then we have a length $\frac{2(q^2-1)}{m}$ $q$-ary  quantum MDS code with minimum distance $d$ equal to any integer in the range $2 \leq d \leq \frac{q-1}{2}+\frac{q+1}{2m}$.}\\

There are many such odd prime powers $q$ as illustrated in the following table.\\

\begin{center}
{\bf Table 6} Quantum MDS codes with lengths $\frac{2(q^2-1)}{m}$\\
\bigskip
\begin{tabular}{||c|c||}\hline
Quantum MDS code& $q, m, d $ \\ \hline
$[[48,48-2d+2,d]]_{13}$&13, 7, $2 \leq d \leq 7$\\ \hline
$[[48, 48-2d+2, d]]_{25}$&17, 9, $2 \leq d \leq 9$\\ \hline
$[[56, 56-2d+2,d]]_{29}$&29, 15, $2 \leq d \leq 15$\\ \hline
$[[144,144-2d+2,d]]_{41}$&37, 19, $2 \leq d \leq 19$ \\ \hline
$[[192, 192-2d+2,d]]_{49}$&49, 25, $2 \leq d \leq 25$ \\ \hline
$[[960, 960-2d+2, d]]_{49}$&49, 5, $2 \leq d \leq 29$ \\ \hline
$[[288, 288-2d+2, d]]_{73}$& 73, 37, $2 \leq d \leq 37$ \\ \hline
$[[1760, 1760-2d+2, d]]_{89}$&89, 9, $2 \leq d \leq 49$ \\ \hline
\end{tabular}
\end{center}

The lengths of some quantum MDS $q$-ary codes in the above table 6  have the form $4(q-1)$ where $q$ is an odd prime power such that $(q+1)$ can not be divisible by $4$.  This case  has not been covered in the previous results (see the table in page 1482 of [3]).\\ 

{\bf Corollary 4.2} {\em If $q$ is an odd prime power of the form $q \equiv 1$ $mod$ $4$, then we have a length $4(q-1)$ $q$-ary quantum MDS code with minimum distance $d$ for each $d$ in the range $2 \leq d \leq \frac{q+1}{2}$}.\\

From the main result in [9] (or see 3 in the table in page 1482 of [3]), only the range $3 \leq d \leq \frac{q-1}{2}$ is allowed. Our result gives quantum $q$-ary MDS $[[4(q-1), 3q-3, \frac{q+1}{2}]]_{q}$ codes when $q=4k+1$ is an odd prime power.\\

{\bf Corollary 4.3} {\em Let  $q$ be an odd prime power. If there exists an even factor  $2(2k+1)$  of $q-1$ such that $4k+1$ is a odd factor of $q+1$. Then we have a length $\frac{q-1}{2k+1} \cdot (q+1)$  $q$-ary quantum MDS codes with any minimum distance $d$ in the range $2 \leq d \leq \frac{q-1}{2}+\frac{q+1}{2(4k+1)}$.}\\

In Theorem 4.11 of [3] and Theorem 2.1 here $m$ cannot be an odd factor. This Corollary 4.3 partially solves this case under a strong assumption on $q$. However there are a lot of such odd prime powers $q$ and even factors $2(2k+1)$ as illustrated in the following table 7.\\

\begin{center}
{\bf Table 7} Quantum MDS codes with lengths $\frac{(q-1)}{2k+1} \cdot (q+1)$\\
\bigskip
\begin{tabular}{||c|c||}\hline
Quantum MDS code& $q, k, d $ \\ \hline
$[[56 \cdot 170, 9520-2d+2, d]]_{169}$&169, 1, $2 \leq d \leq 101$\\ \hline
$[[96 \cdot 290, 27840-2d+2,d]]_{289}$&289, 1, $2 \leq d \leq 173$\\ \hline
$[[456 \cdot 1370,624720-2d+2,d]]_{1369}$&1369, 1, $2 \leq d \leq 821$ \\ \hline
$[[616 \cdot 1850, 1139600-2d+2,d]]_{1849}$&1849, 1, $2 \leq d \leq 1109$ \\ \hline
$[[984 \cdot 6870, 6760080-2d+2, d]]_{6889}$& 6889, 3, $2 \leq d \leq 3709$ \\ \hline
$[[672 \cdot 57122, 38385984-2d+2, d]]_{57121}$&57121, 42, $2 \leq d \leq 28729$ \\ \hline
$[[1896 \cdot 24650, 46736400-2d+2, d]]_{24649}$&24649, 6, $2 \leq d \leq 12817$ \\ \hline

\end{tabular}
\end{center}

\section{New Quantum Codes IV}

In this section we treat the case that $q$ is an odd prime power and $n=\frac{q^2-1}{m}$, where $m|q^2-1$, and $m$ is not a factor of  $q-1$ or $q+1$.\\

We need the following two lemmas.\\

{\bf Lemma 5.1.} {\em If $m_1$ is an even integer and $m_2$ is an odd integer satisfying $gcd(m_1,m_2)=1$, there are infinitely many primes $q$ satisfying $m_1|q-1$ and $m_2|q+1$.}\\

{\bf Proof.} Since $gcd(m_1, m_2)=1$ we have two integers $l_0$ and $k_0$ satisfying $l_0m_1+2=k_0m_2$. Thus $l=l_0+m_2t$ and $k=k_0+m_1t$ also satisfy $lm_1+2=km_2$ for all integers $t=0 \pm1, \pm2, .....$. It is clear $gcd(l_0m_1+1,m_1)=1$. We have $l_0m_1+1+1=k_0m_2$, then $gcd(l_0m_1+1,m_2)=1$. \\

On ther other hand from the famous Dirichlet theorem  there are infinitely many primes in the arithmetic sequence $m_2m_1t+l_0m_1+1$ because of $gcd(l_0m_1+1,m_1m_2)=1$. It is direct to verify $m_1|q-1$ and $m_2|q+1$.\\

{\bf Lemma 5.2.} {\em There are infinitely many pairs of positive integers $(m_1,m_2)$ satisfying the following conditions.\\
1) $m_1$ is even and $m_2$ is odd, $gcd(m_1,m_2)=1$;\\
2) $\frac{m_1+m_2-1}{m_1m_2}=\frac{1}{m}$ where $m$ is a positive integer satisfying $gcd(m_1,m) >1$ and $gcd(m_2,m) >1$.}\\

{\bf Proof.}  We consider $m_2=k_1k_2$ where $k_1$ and $k_2$ are odd numbers. Set $k_3$ and $k_4$ are two un-determined positive integers satisfying $k_1k_2-1+2k_3k_4=k_1k_3$. Then $k_1k_2-1=k_3(k_1-2k_4)$. From the factorization of $k_1k_2-1$ we get suitable $k_3$ and $k_4$.\\

When $k_1=35$ and $k_2=3$, $105-1=8 \cdot 13=k_3(35-2k_4)$. We can set $k_3=8$ and $k_4=11$. Then $m_1=176$ and $m_2=105$. $\frac{105+176-1}{176 \cdot 105}=\frac{1}{66}$. When $k_1=35$ and $k_2=5$. $174=6 \cdot 29=k_3(35-2k_4)$. We can set $k_3=6$ and $k_4=3$. Then $m_1=36$ and $m_2=175$. $\frac{175+36-1}{36 \cdot 175}=\frac{1}{30}$.\\

{\bf Theorem 5.1.} {\em There are infinitely many pairs of integers $(m_1,m_2)$ as in Lemma 5.2 and infinitely many primes $q$ as in Lemma 5.1 for each such pair $(m_1, m_2)$. For each such pair $(m_1,m_2)$ and the infinitely many primes $q$ as in Lemma 5.1, we have a $q$-ary quantum MDS code with length $n=\frac{q^2-1}{m}$ and any minimum distance $d$  in the range $2 \leq d \leq \frac{q-1}{2}+min\{\frac{q+1}{2m_2}, \frac{q-1}{m_1}+1\}$.}\\

{\bf Proof.} The conclusion follows from Lemma 3.1 and Lemma 3.2 and Theorem 2.4 directly. \\

We list some new $q$-ary quantum MDS codes from Theorem 3.1 in the following table.\\

\begin{center}
{\bf Table 8} Quantum MDS codes from Theorem 3.1\\
\bigskip
\begin{tabular}{||c|c||}\hline
Quantum MDS code& $q, m_1, m_2 , d $ \\ \hline
$[[1088 \cdot 1995, 2170560-2d+2, d]]_{11969}$&11969, 176, 105, $2 \leq d \leq 6041$\\ \hline
$[[2768 \cdot 5075, 14047600-2d+2,d]]_{30449}$&30449, 176, 105, $2 \leq d \leq 15369$\\ \hline
$[[7758 \cdot 9310, 72226980-2d+2,d]]_{46549}$&46549, 36, 175, $2 \leq d \leq 23407$ \\ \hline
$[[9858 \cdot 11830, 116620140-2d+2,d]]_{59149}$&59149, 36, 175, $2 \leq d \leq 29743$ \\ \hline
\end{tabular}
\end{center}

{\bf Corollary 5.1.} {\em Let $k$ be any positive integer satisfying  $k \equiv 5$ $mod$ $9$. If $q=16k^2-12k+1$ is an odd prime power  then we have a $q$-ary quantum MDS code with length $\frac{q^2-1}{3k}$ and minimum distance $d$ for each integer $d$ in the range $2 \leq d \leq \frac{q+1}{2}+\frac{2k-1}{3}$.}\\

{\bf Proof.} Set $m_1=4k$ and $m_2=3(4k-1)$ in Theorem 3.1 we get the conclusion.\\

For example when $k=14$ and $q=2969$ is a prime  we have a $2969$-ary quantum MDS $[[209880, 209880-2d+2, d]]_{2969}$ code for each integer $d$ in the range $2 \leq d \leq 1494$. In the above Corollary 5.1 we should note that $3k$ is not a factor of $q-1$ or $q+1$. This case has not been treated in the previous works [3, 8, 9, 10, 11, 12, 13].\\

\section{Summary}

In this paper we give a direct method constructing $q^2$-ary Hermitian orthogonal MDS codes with dimensions $k> \frac{q}{2}$ from  generator matrices. This leads  to many new $q$-ary quantum MDS codes  with minimum distances $d>\frac{q}{2}$. Some new $q$-ary quantum MDS codes with $q> \frac{q}{2}$ constructed in this paper are listed as follows.\\

\newpage

\begin{center}
{\bf Table 9} Quantum MDS codes with minimum distance $d>\frac{q}{2}$\\
\bigskip
\begin{tabular}{||c|c|c|}\hline
Length & Distance & Reference \\ \hline
$n=\frac{q^2-1}{m}$, $m|q+1$, $m$ odd & $2 \leq d \leq \frac{q-1}{2}+\frac{q-1}{2m}$ & [3,21,22]\\ \hline
$n=\frac{q^2-1}{m}$, $m|q+1$, $m$ even &$2\leq d \leq \frac{q-1}{2}+\frac{q-1}{m}$& [21] \\ \hline
$n=\frac{q^2-1}{m}$, $m|q-1$, $m$ even & $2 \leq d \leq \frac{q+1}{2}+\frac{q-1}{m}$ & [3], Theorem 2.1\\ \hline
$n=\frac{q^2+m-1}{m}$, $m|q+1$, $m=2k+1$ odd & $2 \leq d \leq \frac{q+1}{2}+\frac{q-1}{2m}$ & {\bf New}\\ \hline
$n=4(q-1)$, $q \equiv 1$ $mod$ $4$ &$d=\frac{q+1}{2}$& {\bf New}\\ \hline
$n=\frac{2(q^2-1)}{m}$, odd $q$, odd $m|q+1$ \\s.t. $m-1|q-1$ &$ 2 \leq d \leq \frac{q-1}{2}+\frac{q+1}{2m}$& {\bf New} \\ \hline
$n=\frac{q-1}{2k+1} \cdot (q+1)$,  $2k+1|q-1$ \\s.t. $4k+1|q+1$& $2\leq d \leq \frac{q-1}{2}+\frac{q+1}{2(4k+1)}$ & {\bf New} \\ \hline
$n=\frac{(m_1+m_2-1)(q^2-1)}{2m_1m_2}$, \\odd $m_1<m_2$, $gcd(m_1,m_2)=1$, \\$q=2m_1m_2+1$ &$2 \leq d \leq \frac{q+1}{2}+m_1$&{\bf New}\\ \hline
$n=\frac{(m_1+m_2-1)(q^2-1)}{2m_1m_2}$, \\odd $m_1=2k_1+1<m_2=2k_2+1$, \\$gcd(m_1,m_2)=1$ &$2 \leq d \leq \frac{q+1}{2}+\frac{q-1}{2(2k_2+1)}$&{\bf New}\\ \hline
$n=\frac{m_1m_2+m_2m_3+m_1m_3-m_1-m_2-m_3+1(q^2-1)}{2m_1m_2m_3}$, \\odd $q=2m_1m_2m_3+1$, odd $m_1<m_2<m_3$, \\ $gcd(m_{i_1},m_{i_2})=1$ & $2 \leq d \leq \frac{q+1}{2}+m_1m_2$ & {\bf New} \\ \hline
$n=\frac{q^2-1}{m}$, suitable $q$ and $m$ not dividing $q-1$\\ or $q+1$, $A \geq 1$  as in Theorem 3.1& $2 \leq d \leq \frac{q-1}{2}+A$& {\bf New} \\ \hline
$n=\frac{q^2-1}{3k}$, $k\equiv 5$ $mod$ $9$,  for odd prime power \\$q=16k^2-12k+1$& $2 \leq d \leq \frac{q+1}{2}+\frac{2k-1}{3}$ & {\bf New} \\ \hline
\end{tabular}
\end{center}

\begin{center}
References
\end{center}

[1] S. A. Aly, A. Klappenecker and P. K. Sarvepalli, On quantum and classical BCH codes, IEEE Trans. Inf. Theory, vol. 53, no. 3, pp. 1183-1188, Mar. 2007.

[2] A. Ashikhmin and E. Knill, Nonbinary quantum stablizer codes, IEEE Trans. Inf. Theory, vol. 47, no. 7, pp. 3065-3072, Nov. 2001.

[3] Bocong Chen, San Ling and Guanghui Zhang, Application of constacyclic codes to quantum MDS codes, IEEE Transactions on Information Theory, vol.61(2015), no.3, 1474-1484.

[4] K. Feng, Quantum code $[[6, 2, 3]]_ p$ and $[[7, 3, 3]]_p$ ( $p \geq 3$) exists, IEEE Trans. Inf. Theory, vol. 48, no. 8, pp. 2384-2391, Jan. 2002.

[5] M. Grassl, T. Beth, and M. Roetteler, On optimal quantum codes, Int. J. Quantum Inform., vol. 2, no. 1, pp. 757-766, 2004.

[6] M. Grassl, M. Roetteler, and T. Beth, On quantum MDS codes, In Proc. Int. Symp. Inform. Theory, Chicago, USA, p.356, 2004.

[7] M. Grassl, M. Roetteler, Quantum MDS codes over small fields, arXiv 1502:05267.

[8] G. G. La Guardia, New  quantum MDS codes, IEEE Transactions on Information Theory, vol.57, no.8, pp.5551-5554. 2011

[9] L. Jin, S. Ling, J. Luo, and C. Xing, Application of classical Hermitian self-orthogonal MDS codes to quantum MDS codes, IEEE Trans. Inf. Theory, vol. 56, no. 9, pp. 4735-4740, Sep. 2010.

[10] L. Jin and C. Xing, Euclidean and Hermitian self-orthogonal algebraic geometry codes and their Application to quantum codes, IEEE Trans. Inform. Theory, vol. 58, pp. 5484-5489, 2012.

[11] L. Jin and C. Xing, A Construction of New Quantum MDS Codes, IEEE Trans. Inform. Theory, vol.60, no. 5, 2921-2925, 2014.

[12] X. Kai and S. Zhu, New quantum MDS codes from negacyclic codes, IEEE Trans. Inform. Theory, vol. 59, no. 2, pp. 1193-1197, Feb. 2013.

[13] X. Kai, S. Zhu, and P. Li, Constacyclic codes and some new quantum MDS codes, IEEE Trans. on Inf. Theory, vol60, no.4, pp.2080-2086, 2014.

[14] E. Knill and R. Laflamme, Theory of quantum error-correcting codes, Phys. Rev. A, vol. 55, no. 2, pp. 900-911, 1997.

[15] R. Laflamme, C. Miquel, J. P. Paz, and W. H. Zurek, Perfect Quantum Error Correcting Code, Phys. Rev. Lett., vol. 77, no. 1, pp. 198-201, July 1996.

[16] Z. Li, L. J. Xing, and X. M. Wang, Quantum generalized Reed-Solomon
codes: Unified framework for quantum MDS codes, Phys. Rev. A,
vol. 77, no. 1, pp. 012308-1-12308-4, 2008.

[17] R. Li and Z. Xu, Construction of $[[n, n-4, 3]]q$ quantum MDS
codes for odd prime power q , Phys. Rev. A, vol. 82, no. 5,
pp. 052316-1-052316-4, 2010.

[18] F. J.  MacWilliams and N. J. A. Sloane, Theory of Error-Correcting Codes, North Holland, Amsterdam, 2nd printing, 1978.

[19] P. W. Shor, Scheme for reducing decoherence in quantum computer
memory, Phys. Rev. A, vol. 52, no. 4, pp. R2493-R2496, 1995.

[20] A. M. Steane, Enlargement of Calderbank-Shor-Steane quantum codes, IEEE. Trans. Inf. Theory, vol. 45, no. 7, pp. 2492-2495,
Nov. 1999.

[21] Liqi Wang and Shixin Zhu, New quantum MDS codes derived from constacyclic codes, arXiv 1405:5421v1, Quantum Information Processing, vol.14, no.3, pp.881-889, 2015.\\

{\bf Xianmang He} was born in Zhejiang Province of China in 1981. He obtained his Ph. D  in the  School of Computer Science of Fudan Universiy in 2011. He is currently a lecturer in the School of Information Sciences and Technology of Ningbo University. His research interests are coding, cryptography and data base security.\\

{\bf Liqing Xu} was born in Shanghai in 1964. She obtained her Master degree in applied mathematics in Fudan University in 1991. She is currently a lecturer in the School of Sciences of Hangzhou Dianzi University. Her main research fields are coding and cryptography.\\

{\bf Hao Chen} was born in Anhui Province of China in 1964. He obtained his Ph.D in Mathematics in Fudan University in 1991. He is currently a professor in the Department of 
Mathematics, School of Sciences of Hangzhou Dianzi University, Zhejiang Province,
China. His research interests are cryptography and coding, quantum
information and computation, algebraic geometry and lattices.

\end{document}